\documentclass[preprint,aps,showpacs,amsfonts,epsf]{revtex4}

\usepackage{graphicx}

\newcommand{\s}{\sigma}

\newcommand{\be}{\begin{equation}}
\newcommand{\ee}{\end{equation}}
\newcommand{\bea}{\begin{eqnarray}}
\newcommand{\eea}{\end{eqnarray}}
\newcommand{\ba}{\begin{array}}
\newcommand{\ea}{\end{array}}
\def\J#1#2#3#4{{#1} {\bf #2}, #3 (#4)}
\def\PRD{Phys. Rev. D}

\def\PRL{Phys. Rev. Lett.}

\def\JMP{J. Math. Phys.}

\def\CQG{Class. Quantum Grav.}
\def\CMP{Commun. Math. Phys.}

\def\PTEP{Prog. Theor. Exper. Phys.}

\begin{document}
\draft
\title{On non-disk geometry of $r=0$ in Kerr-de Sitter\\ and Kerr-Newman-de Sitter spacetimes}

\author{V.~S.~Manko and H. Garc\'ia-Compe\'an}
\address{Departamento de F\'\i sica, Centro de Investigaci\'on y de
Estudios Avanzados del IPN, A.P. 14-740, 07000 M\'exico D.F.,
Mexico}

\begin{abstract}
Gaussian curvature of the two-surface $r=0$, $t={\rm const}$ is
calculated for the Kerr-de Sitter and Kerr-Newman-de Sitter
solutions, yielding non-zero analytical expressions for both the
cases. The results obtained, on the one hand, exclude the
possibility for that surface to be a disk and, on the other hand,
permit one to establish a correct geometrical interpretation of
that surface for each of the two solutions.
\end{abstract}


\maketitle

%\twocolumn

\section{Introduction}

It is well known that the two-surface $r=0$, $t={\rm const}$ of
the Kerr solution \cite{Ker} has zero Gaussian curvature
\cite{ONe1} and therefore is commonly interpreted as a disk
\cite{BLi,HEl} or a ``flat sphere'' composed of two disks of
radius $a$ joined together along the ring singularity~\cite{ONe2}.
For not quite perspicuous reasons, these interpretations had been
automatically extrapolated to the analogous surfaces in
generalized Kerr black-hole spacetimes involving electromagnetic
field or cosmological constant, as may be readily inferred by
examining for instance the Penrose-Carter conformal diagrams of
the Kerr-Newman (KN) or Kerr-de Sitter (KdS)
spacetimes~\cite{NCC,Car1,GHa} where the regions of positive and
negative radial coordinate $r$ are supposedly glued on such disks.
Recently, however, using a direct calculation \cite{GCM}, the
Gaussian curvature of the surface $r=0$, $t={\rm const}$ in the KN
case has been shown to be a function of the polar coordinate
$\theta$, thus clearly disproving the disk interpretation of the
latter surface in that case. Moreover, a study of the above
two-surface in the Weyl-Papapetrou cylindrical coordinates carried
out in the same paper \cite{GCM} has resulted in a novel
interpretation of that surface even in the case of the Kerr
solution -- a {\it dicone} instead of a disk -- which looks
plausible since a dicone is a closed surface having, like a disk,
zero Gaussian curvature and in addition better fitting the
corresponding surface's equation in cylindrical coordinates.

The present communication is aiming, firstly, to provide
convincing evidence against the disk interpretation of the surface
$r=0$, $t={\rm const}$ in the case of two well-known black-hole
solutions with a non-zero cosmological constant, namely, the KdS
and Kerr-Newman-de Sitter (KNdS) spacetimes~\cite{Car1,GHa,Car2},
and, secondly, to establish a correct geometrical interpretation
of that important surface and briefly discuss the main
mathematical and physical implications engendered by the new
geometry.

\section{The surface $r=0$, $t={\rm const}$ in KdS spacetime}

The KdS metric was obtained by Carter, and in the
Boyer-Lindquist-like coordinates it has the form
\be d s^2=\Sigma\left(\frac{d r^2}{\Delta_r}
+\frac{d\theta^2}{\Delta_\theta}\right)
+\frac{\Delta_\theta\sin^2\theta}{\Sigma}[a d
t-(r^2+a^2)d\varphi]^2-\frac{\Delta_r}{\Sigma}(d t-a\sin^2\theta
d\varphi)^2, \label{KdS_m} \ee
where
\bea \Delta_r&=&(r^2+a^2)\left(1-\frac{\Lambda}{3}r^2\right)-2Mr,
\quad \Delta_\theta=1
+\frac{\Lambda}{3}a^2\cos^2\theta, \nonumber\\
\Sigma&=&r^2+a^2\cos^2\theta, \label{dd} \eea
the parameters $M$ and $a$ being related to the mass and angular
momentum per unit mass of the source, and $\Lambda$ being the
cosmological constant. The ring singularity of this spacetime
corresponds to $r=0$, $\theta=\pi/2$.

Since our interest lies basically in establishing the geometry of
the surface $r=0$, $t={\rm const}$, we are not going to touch here
the general properties of the KdS solution, restricting ourselves
to only mentioning that their discussion can be found, e.g., in
the papers \cite{GHa,AMa} and monograph \cite{GPo}. With that
said, we go directly to the two-surface we are interested in and
write out its line element,
\be d\s^2=\frac{a^2\cos^2\theta}{\Delta_\theta}d\theta^2
+a^2\left(1+\frac{\Lambda}{3}a^2\right)\sin^2\theta d\varphi^2,
\label{KdS2} \ee
which does not contain the mass parameter $M$.

To calculate the Gaussian curvature $K$ of the surface
(\ref{KdS2}) which equals half the scalar curvature $R$, we have
used an utterly user-friendly computer program ``Ricci''
\cite{Agu} and obtained the following very simple formula
\be K=\frac{\Lambda}{3}. \label{G1} \ee
Though it is tempting to conclude from (\ref{G1}) that the surface
(\ref{KdS2}) is a sphere or a pseudosphere depending on whether
$\Lambda>0$ or $\Lambda<0$, such a conclusion would not be really
correct since, for instance, Liebmann's theorem on the closed
surfaces with positive $K$ is applicable to {\it regular} surfaces
only. At the same time, as it follows from (\ref{KdS2}), our
two-surface has a singularity at $\theta=\pi/2$ (it is the ring
singularity of KdS solution), and it may also be non-regular at
the points $\theta=0$ and $\theta=\pi$. It is clear as well that
in the $\Lambda<0$ case the surface of negative Gaussian curvature
cannot be a tractricoid because the latter then would have been
stretching along the entire symmetry axis. Taking into account
that the $r=0$, $t={\rm const}$ surface of the Kerr solution
($\Lambda=0$) is a dicone depicted in Fig.~1, it would be
plausible to suppose that in the Kerr-anti-de Sitter case the
respective surface is represented by some conic surface of a
constant negative Gaussian curvature with singular vertices and
equator, like the one shown in Fig.~2. By analogy, the surface
$r=0$, $t={\rm const}$ of the KdS solution with $\Lambda>0$ is
represented by a conic surface of rotation with constant positive
Gaussian curvature and singular vertices and equator (see Fig.~3).
For the theory of such surfaces and practical aspects of their
construction we refer the reader to the monograph \cite{Gra}.

The surfaces from figures 2 and 3, which we shall dub,
respectively, {\it a concave dicone} and {\it a convex dicone of
constant Gaussian curvature}, are remarkable in several regards.
First of all, and most importantly, they leave no doubt that the
surface $r=0$, $t={\rm const}$ of the KdS solution {\it is not a
disk}. Moreover, they provide a strong support to the novel
interpretation of the above surface in the Kerr solution -- a
dicone of zero Gaussian curvature -- which may be considered as
the simplest ($\Lambda=0$) non-trivial specialization of the
general KdS case. It is also surprising that the cosmological
constant $\Lambda$ in the KdS metric modifies the surface $r=0$,
$t={\rm const}$ of the pure Kerr spacetime in such a way that the
Gauss curvature of that surface in the presence of $\Lambda$ still
remains constant, something that does not happen, as will be seen
in the next section, when a charge parameter is introduced into
the solution.

\section{The surface $r=0$, $t={\rm const}$ in KNdS spacetime}

The charged version of the KdS field, the KNdS solution, was also
obtained by Carter, but its conventional form currently used in
the literature is due to the paper of Gibbons and
Hawking~\cite{GHa}; the solution is determined by the line element
\be d s^2=\Sigma\left(\frac{d r^2}{\Delta_r}
+\frac{d\theta^2}{\Delta_\theta}\right)
+\frac{\Delta_\theta\sin^2\theta}{\Xi^2\Sigma}[a d
t-(r^2+a^2)d\varphi]^2-\frac{\Delta_r}{\Xi^2\Sigma}(d
t-a\sin^2\theta d\varphi^2, \label{KNdS_m} \ee
where
\bea
\Delta_r&=&(r^2+a^2)\left(1-\frac{\Lambda}{3}r^2\right)-2Mr+Q^2,
\quad \Delta_\theta=1
+\frac{\Lambda}{3}a^2\cos^2\theta, \nonumber\\
\Sigma&=&r^2+a^2\cos^2\theta, \quad \Xi=1+\frac{\Lambda}{3}a^2
\label{dd2}, \eea
and $Q$ is the charge parameter. The associated electromagnetic
four-potential $A_i$ is given by
\be
A_i=-\frac{Qr}{\Xi\Sigma}(\delta_i^t-a\sin^2\theta\delta_i^\varphi).
\label{Ai} \ee

Compared to the metric (\ref{KdS_m}), the line element
(\ref{KNdS_m}) contains the factor $\Xi$ appearing as a result of
rescaling the coordinates $\varphi$ and $t$; however, as will be
seen below, this factor does not affect anyhow the intrinsic
geometry of the surface $r=0$, $t={\rm const}$.

The latter two-surface is defined by the line element
\be d\s^2=\frac{a^2\cos^2\theta}{\Delta_\theta}d\theta^2
+(a^2\Xi-Q^2\tan^2\theta)\sin^2\theta d\varphi^2, \label{KNdS2}
\ee
and the corresponding Gaussian curvature can be shown to have the
form
\bea K&=&\frac{\sec^4\theta}{3a^2(a^2\Xi\cos^2\theta
-Q^2\sin^2\theta)^2} \{\Lambda a^6\Xi^2\cos^8\theta \nonumber\\
&&+a^2Q^2\Xi\cos^2\theta[9+3\sin^2\theta+\frac{\Lambda}{4}
a^2\cos^2\theta(11+\cos4\theta)] \nonumber\\
&&-Q^4\sin^2\theta[6+3\sin^2\theta+\frac{\Lambda}{8}
a^2\cos^2\theta(15+\cos4\theta)]\}. \label{G2} \eea

By setting $Q=0$ in (\ref{G2}), one recovers the constant value
(\ref{G1}) of the KdS solution, thus demonstrating that the
rescaling factor $\Xi$ does not modify the Gaussian curvature of
the surface under consideration, and in the limit $\Lambda=0$ one
arrives at the respective $K$ of the KN space obtained in
\cite{GCM}:
\be K=\frac{Q^2[3a^2-(a^2+Q^2)\sin^2\theta(2+\sin^2\theta)]}
{a^2\cos^4\theta(a^2\cos^2\theta-Q^2\sin^2\theta)^2}. \label{GKN}
\ee

Comparison of formulas (\ref{G1}), (\ref{G2}) and (\ref{GKN})
leads to a conclusion that the effect of the charge parameter on
the geometry of the surface $r=0$, $t={\rm const}$ of the Kerr
solution is more significant than the analogous effect produced by
the cosmological constant $\Lambda$, and also that the combined
effect of the parameters $Q$ and $\Lambda$ distorts drastically
that surface. It has already been shown in \cite{GCM} that formula
(\ref{GKN}) defines a specific surface of revolution of positive
and negative Gaussian curvature; in this respect, the surface
determined by (\ref{G2}) must in fact be given exactly the same
interpretation as in the KN case, with the only additional remark
that obviously the particular form of the regions of positive and
negative Gaussian curvature defined by zeros of the numerator in
(\ref{G2}) cannot be studied analytically, needing a numerical
analysis. Therefore, one can see that the surface $r=0$, $t={\rm
const}$ of the KNdS solution cannot be interpreted as a disk,
rather being a surface of revolution with various regions of
positive and negative Gaussian curvature dependent on the polar
coordinate $\theta$. We leave to the reader's imagination the task
of deforming (in the equatorially symmetric manner) the dicones
from Figs.~2 and 3 for having an idea of how that surface may look
like for positive and negative values of $\Lambda$.

\section{Discussion}

The established fact that the surface $r=0$, $t={\rm const}$ in
the KdS and KNdS solutions is not a disk has several mathematical
and physical consequences. First of all, it is now obvious that,
if $r\ge 0$, the usual Boyer-Lindquist-like coordinates do not
cover the whole KdS and KNdS manifolds, leaving the interior
regions of dicones beyond their attainability. In this respect,
the above surface of a stationary black hole cannot be considered
as its center, contrary to what was suggested in \cite{ONe2}, as
it is apparent that the black hole's center must coincide with the
geometrical center of the ring singularity, which is also the
geometrical center of a concrete dicone containing that
singularity.

The non-disk geometry of the surface $r=0$, $t={\rm const}$ seems
to invalidate the known approaches to the extensions of the
black-hole rotating spacetimes to infinite negative values of the
radial coordinate $r$ requiring an artificial gluing of the spaces
with $r\ge0$ and $r\le0$ on the ``disks''. Indeed, the supposition
that this surface is a disk (in a non-extended spacetime, to avoid
confusion) implies that the disk lies in the equatorial plane, so
that by crossing it from the upper hemisphere
($0\le\theta\le\pi/2$), one immediately finds oneself in the other
hemisphere ($\pi/2<\theta\le\pi$). However, if the above surface
is represented by any of the dicones considered in the previous
section, then by crossing a dicone at some $0<\theta<\pi/2$, one
still will be staying in the same hemisphere, apparently needing
to cover some distance in order to reach the equatorial plane; by
crossing the latter, one will find oneself in the second
hemisphere still inside the dicone, and will need to run another
way for being able to eventually go out of the dicone at some
$\pi/2<\theta<\pi$. The interior of the dicone can be appended to
the general manifold corresponding to $r\ge0$ in various ways, for
instance by redefining the radial coordinate $r$ in the manner
suggested in \cite{GCM} for the KN solution, which would be
equivalent to extending $r$ to a {\it finite} negative value
$r_0<0$ (the possibility discussed in \cite{MRu}), or by
introducing an appropriate set of coordinates fully covering the
regions exterior and interior to the dicone's surface. Concerning
the latter possibility it is worth remarking that the introduction
of a new coordinate set may in principle depend on the sign of the
mass parameter $M$, like it takes place, e.g., in the case of the
KN solution whose exterior and interior regions are rather well
described by the usual Weyl-Papapetrou cylindrical coordinates
only when $M<0$ \cite{GCM,MRu}. Incidentally, in the presence of a
non-zero cosmological constant $\Lambda$ the problem of
introducing the Weyl-Papapetrou-like coordinates in the stationary
axisymmetric solutions is more complicated than in the $\Lambda=0$
case and is likely to be studied in more detail in the future (an
important work in this direction has been recently done
in~\cite{CLS,Ast}).

Though it is clear that the surface $r=0$, $t={\rm const}$ is not
smooth and therefore an analytic extension through it may exhibit
problems of differentiability (that is why a set of coordinates
better than that of Boyer and Lindquist is actually needed), one
still could ask oneself a question about an exact negative value
of the aforementioned finite $r_0$ at which the center of the ring
singularity will be attained during the extension procedure
through the dicone's surface by using directly the Boyer-Lindquist
coordinates. This question is not as trivial as may look like, and
at the moment we cannot give an exhaustive answer to it. At the
same time, taking into account that the Gauss curvature of a
sphere is $K=1/r^2$, being singular at $r=0$, it might look
plausible as a possibility to associate the center of the ring
singularity of the KdS and KNdS black-hole solutions with one of
the singularities of the Gaussian curvature of the corresponding
surface $r={\rm const}$, $t={\rm const}$. While the calculation of
$K$ for the latter surface does not represent technical
difficulties, the resulting expression, however, is very
cumbersome and we do not give it here. Instead, we write out below
the form of that $K$ in the particular $\theta=\pi/2$ case when
the general expression considerably simplifies, yielding
\be K=\frac{\Xi r^4(r^2+a^2)+a^2(2r^2+a^2)(2Mr-Q^2)} {r^4[\Xi
r^2(r^2+a^2)+a^2(2Mr-Q^2)]}. \label{Ke} \ee
The above formula describes Gaussian curvatures of the surfaces
$r={\rm const}$, $t={\rm const}$ of the KNdS solution at the
points located in the equatorial plane, and the apparent
singularity at $r=0$ corresponds to the usual ring singularity
$r=0$, $\theta=\pi/2$. Therefore, starting from some positive $r$
and moving in the equatorial plane towards the center, one first
comes to the singular point $r=0$ and then, after passing it,
finds oneself in the zone of negative $r$ inside the region
enclosed by the surface $r=0$, $t={\rm const}$. The second factor
in the denominator of (\ref{Ke}) is a quartic polynomial in $r$;
then, if we set for simplicity $Q=0$, this factor further
factorizes into $r$ and $\Xi r(r^2+a^2)+2Ma^2$, the latter having
the following real negative root $r_0$ when $\Xi>0$, $M>0$:
\be r_0=\frac{a^{2/3}(x_0^{2/3}-a^{2/3}\Xi^{2/3})} {\sqrt{3}\,(\Xi
x_0)^{1/3}}, \quad x_0=\sqrt{27M^2+a^2\Xi^2} -3\sqrt{3}\,M.
\label{r0} \ee
In figure 4 we have plotted a characteristic dependence of $r_0$
on $a$ for the particular value of the mass parameter $M=2$ and
three different values of $\Lambda$: $-1, 0, +1$ (these correspond
to dashed, solid and dotted lines, respectively), whence it
follows that $|r_0|$ is an increasing function of $a$, and also
that for a given value of $a$ the respective $|r_0|$ is the
largest in the case of negative $\Lambda$. At the same time, the
question whether the above $r_0$ really represents the center of
the ring singularity still may require further clarification.

The main physical implication of our results consists in giving a
fairly new picture of the internal structure of rotating black
holes in the vicinity of the ring singularity. As a matter of
fact, it is clear now that in an appropriate coordinate atlas
fully covering the regions exterior and interior to the closed
conic surfaces of revolution the ring singularity will be smoothly
traversable by the observers who may cross it from one hemisphere
to another in any direction as many times as they like, always
staying in the same black-hole spacetime -- no introduction of an
additional copy of the solution with another spatial infinity for
artificially attaching it to the non-smooth surface $r=0$, $t={\rm
const}$ is required.

\section*{Acknowledgments}

This work was partially supported by Project~128761 from CONACyT
of Mexico.

\newpage

\begin{figure}
  \centering
    \includegraphics[width=70mm]{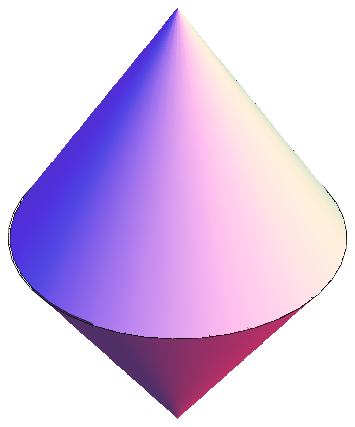}
  \caption{The two-surface $r=0$, $t={\rm const}$ of
  the Kerr solution -- a dicone of zero Gaussian curvature.}
  \label{fig1}
\end{figure}

\begin{figure}
  \centering
    \includegraphics[width=80mm]{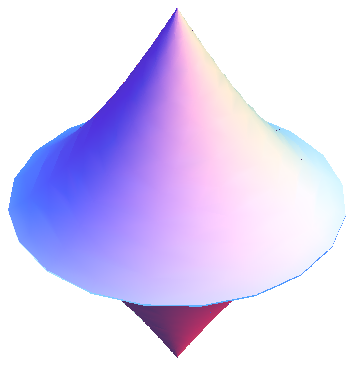}
  \caption{The two-surface $r=0$, $t={\rm const}$ of
  the KdS solution with $\Lambda<0$ -- a conic surface of revolution of
  constant negative Gaussian curvature.}
  \label{fig2}
\end{figure}

\begin{figure}
  \centering
    \includegraphics[width=60mm]{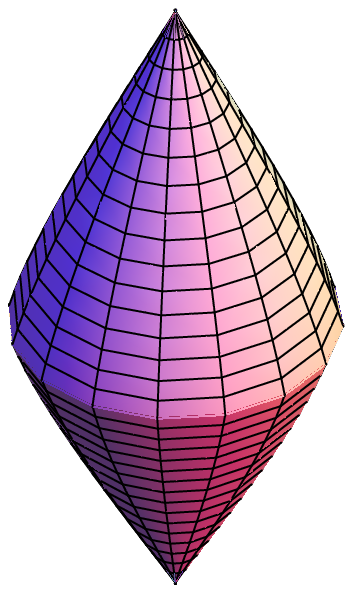}
  \caption{The two-surface $r=0$, $t={\rm const}$ of
  the KdS solution with $\Lambda>0$ -- a conic surface of revolution of
  constant positive Gaussian curvature.}
  \label{fig2}
\end{figure}

\begin{figure}
  \centering
    \includegraphics[width=80mm]{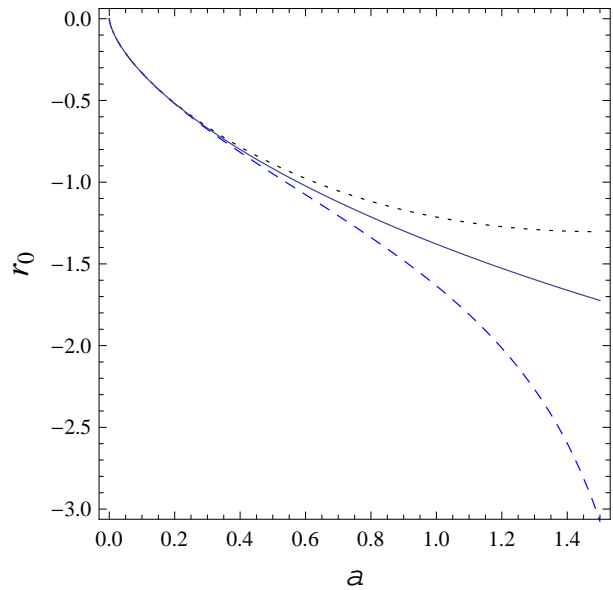}
  \caption{The $r_0$ versus $a$ plots for $M=2$ and three particular
   values of $\Lambda$: $-1$ (dashed line), 0 (solid line) and $+1$ (dotted line).}
  \label{fig2}
\end{figure}

\end{document}